\documentclass[11pt,twoside]{article}
\usepackage{./asp2014}

\aspSuppressVolSlug
\resetcounters

\begin{document}

\title{HR\,62: A New Evolved Chemically Peculiar Late-B Star?}
\author{Tolgahan~K{\i}l{\i}\c{c}o\u{g}lu$^1$ and Richard~Monier$^2$
\affil{$^1$Ankara University, Faculty of Science, Department of Astronomy and Space Sciences, D\"{o}gol Cad., 06100, Yenimahalle, Ankara, Turkey; \email{tkilicoglu@ankara.edu.tr}}
\affil{$^2$LESIA, UMR 8109, Observatoire de Paris et Universit\'{e} Pierre et Marie
Curie Sorbonne Universit\'{e}s, place J. Janssen, Meudon, Paris, France; \email{Richard.Monier@obspm.fr}}
}

% This section is for ADS Processing.  There must be one line per author.
\paperauthor{Tolgahan~K{\i}l{\i}\c{c}o\u{g}lu}{tkilicoglu@ankara.edu.tr}{0000-0001-6947-822X}{Ankara University Faculty of Science}{Department of Astronomy and Space Sciences}{Ankara}{}{06100}{Turkey}
\paperauthor{Richard~Monier}{Richard.Monier@obspm.fr}{0000-0002-0735-5512}{LESIA Observatoire de Paris et Universit\'{e} Pierre et Marie Curie Sorbonne Universit\'{e}s}{Physics Department}{Paris}{}{UMR 8109}{France}

\begin{abstract}
The spectrum of the evolved B8\,III giant HR\,62 exhibits weak He-lines and strong Mn and P lines. HR\,62 therefore resembles both a HgMn star (CP3) and a He-weak\,PGa star (CP4).

This study is a companion project to a high resolution survey of slowly rotationg late-B type stars aiming at finding new chemically peculiar stars. We have analysed the spectra of HR\,62 (B8\,III) and the dwarf comparison star HR\,677 (B8\,V) to derive their chemical abundances. Both stars have similar effective temperatures (12500 K) and projected rotational velocities ($\sim$25 km\,s$^{-1}$).

The medium resolution ($R$\,$\sim$14000) spectra covering the wavelength range of 4380-7350 \AA{} of HR\,62 and HR\,677 have been obtained with the \'{e}chelle spectrograph attached to the 40 cm telescope in Ankara University Kreiken Observatory (AUKR), Turkey.  We have used SYNSPEC49/SYNPLOT written by I. Hubeny, T. Lanz to compute grids of synthetic spectra and derive elemental abundances by modeling selected unblended lines.

We find that HR\,62 exhibits underabundance of Si and remarkable overabundance of P and Mn with respect to the Sun. In contrast, HR\,677 does not have abundances departing by more than $\pm$ 0.25 dex from solar abundances. A mass of 5.4 $M_\odot$ and an age of 90 Myr have been estimated for HR\,62.

We discuss the origin of the chemical peculiarities of HR 62 and its status as a CP star. The effective temperature of the star (12500 K) agrees well with those of HgMn (CP3) stars. Furthermore, the main sequence end of its evolutionary track also intersects the domain of He-weak\,CP4 stars. Hence these first results suggest that HR\,62 may be a transition object between the CP4 to CP3 subgroup. However, a more detailed analysis of higher resolution spectra at shorter wavelengths (< 4380 \AA{}) is necessary to clearly address the nature of this interesting object.

\end{abstract}

\section{Introduction}

The causes of the chemical peculiarities in the atmospheres of non-magnetic B and A type stars have not yet been understood fully. Atomic diffusion is one of the most likely explanations \citep{michaudetal15}. In order to test the predictions of evolutionary models including atomic diffusion, elemental abundances of a large number of CP stars should be derived. 

\articlefigure{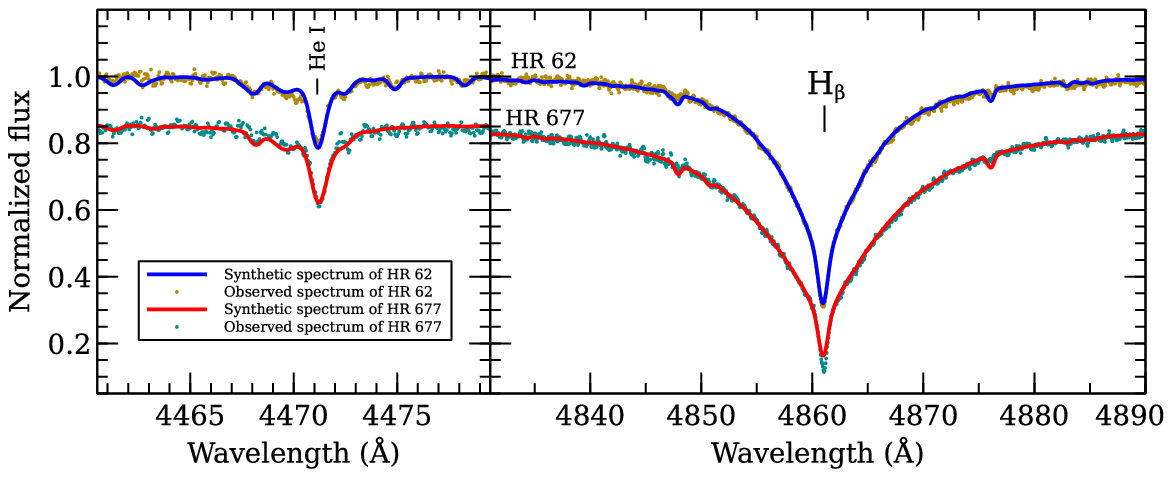}{fig1}{Modelling of the $H_{\beta}$ and He I lines in the spectra of HR\,62 and HR\,677 (the spectrum of HR\,677 is shifted by -0.15 in y-axis).}

We have selected about 100 slowly rotating bright ($V < 7^{\rm{m}}$) late-B stars in order to look for new CP stars. A first step in this project consists in observing these objects using a 40-cm telescope equiped with a medium resolution \'{e}chelle spectrograph to detect new CP stars. The CP candidates will then be observed at a higher resolution with larger telescopes. During this survey, we have found strong Mn {\small II} and P {\small II} lines in the spectrum of HR\,62. This study presents the elemental abundance analysis for HR\,62 and the dwarf comparison star HR\,677 based on their medium resolution spectra. 

\section{Observations and Analysis}

The medium resolution spectra (R$\sim14000$) of HR 62 and HR 677 covering the range 4380-7350\,\AA\ were obtained using the Shelyak eShel spectrograph mounted on the 40-cm telescope in Ankara University Kreiken Observatory (AUKR) in 2017. The model atmospheres were computed using {\small ATLAS12} \citep{sbordoneetal04,kurucz05} assuming local-thermodynamic, radiative and hydrostatic equilibria. Grids of synthetic spectra were computed using {\small SYNSPEC49/SYNPLOT} \citep{hubenylanz95}. The linelist was first constructed from R. Kurucz's gfall.dat and then updated using {\small VALD} and {\small NIST} atomic database.

The effective temperatures of the stars were initially estimated from their spectral types and their Johnson $BV$ magnitudes using the calibrations of \citet{flower96} and \citet{besselletal98}. The color excess $E(B-V)$ of the stars were retrieved from 3D dust map of \citet{greenetal18}. The same effective temperature ($12500\pm500$ K) was found for both stars. The surface gravity of the stars were derived by modelling $H_\mathrm{\beta}$ lines in the spectra (Fig. \ref{fig1}). The logarithms of the surface gravities are $3.20\,\pm\,0.05$ and $3.95\,\pm\,0.05$ (gravity $g$ in $\rm{cm\,s^{-2}}$)  for HR\,62 and HR\,677, respectively.

\articlefigure{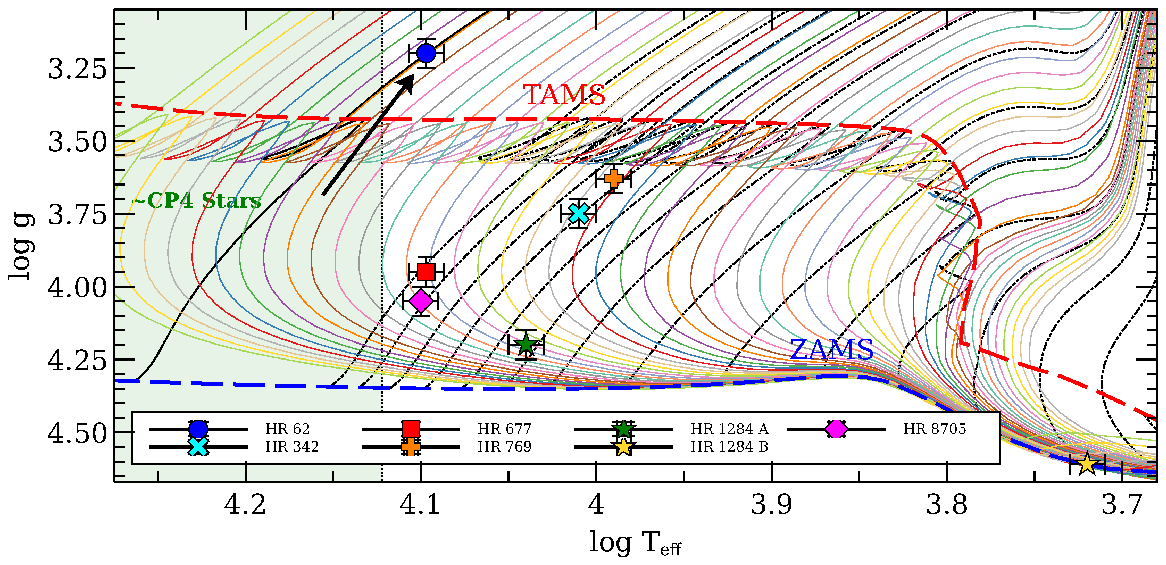}{fig2}{Position of HR\,62 and HR\,677 on log\,$g$--$T_\mathrm{eff}$ diagram with theoretical evolutionary models and isochrones. The other stars analyzed up to now in the same project were also taken from \citet{kilicoglu18} and plotted for comparison.}

\articlefigure{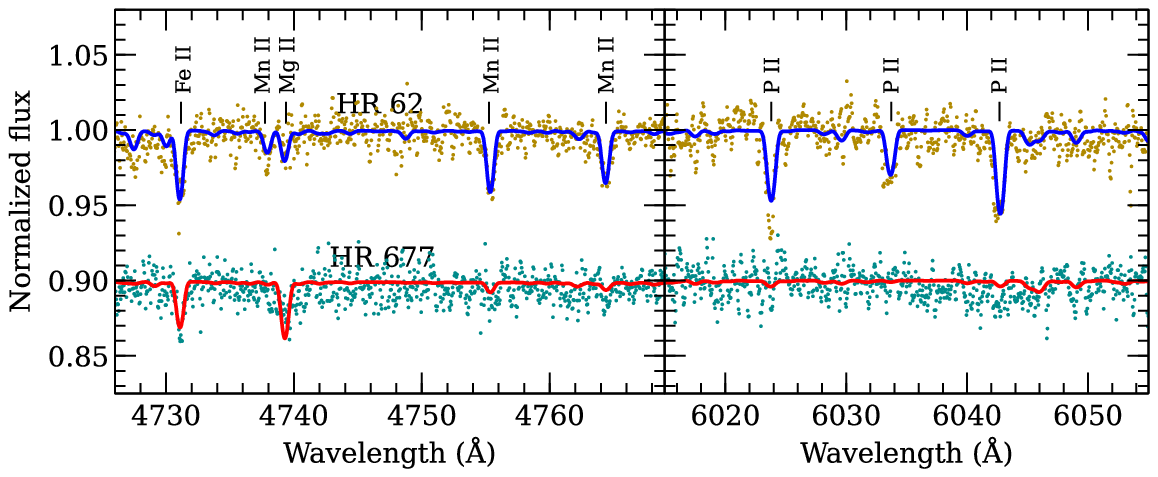}{fig3}{Same in Fig. \ref{fig1}, but for Fe II, Mg II, Mn II, and P II lines.}

The masses and ages of the stars were also estimated by placing them in a \mbox{log\,$g$--$T_\mathrm{eff}$}  diagram (Fig. \ref{fig2}). The theoretical evolutionary tracks and isochrones were taken from \citet{bressanetal12}. Masses of 5.40 and 3.55 $M_\odot$ and ages of 90 and 173 Myr were found for HR\,62 and HR\,677, respectively. We iteratively adjusted the synthetic spectra to the observed unblended lines until the best fit was achieved in order to derive elemental abundances. We used a Levenberg-Marquardt algorithm \citep{markwardt09} for the $\chi^{2}$ minimisation. Fig. 3 illustrates the modeling of Fe, Mn, Mg, and P lines in two spectral regions. Clearly Mn and P are more abundant in the atmosphere of HR\,62 than in HR\,677.

\articlefigure{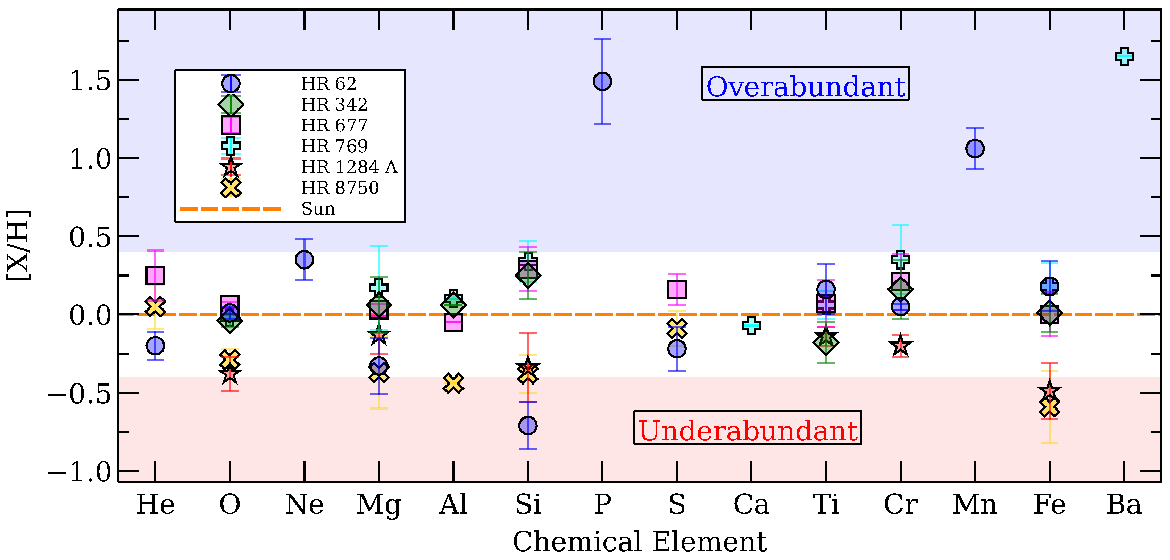}{fig4}{Chemical abundance pattern of the late-B type stars observed in AUKR.}
 
\section{Results and Discussion}

We have derived the abundances of 11 elements for HR\,62 and 9 elements for HR\,677.  Fig. \ref{fig4} shows the abundance patterns of HR\,62, HR\,677 and the other stars analyzed so far using the test observations of the 40-cm telescope and medium resolution \'{e}chelle spectrograph in AUKR. Using the solar abundances of \citet{grevessesauval98}, we find that the abundances of Mn and P appear to be remarkably enhanced in the photosphere of HR\,62. This star also exhibits a slight overabundance of Ne, slight underabundance of He and Mg, and underabundance of Si. In contrast, the abundances of HR\,677  do not depart significantly from the adopted solar composition. We cannot yet confirm the presence of strong Ga {\small II} and Hg {\small II} lines in the spectra of HR\,62 as the spectra do not extend to wavelengths shorter than 4380\,\AA{}. The abundance pattern of the star, however, suggest that it is most likely a CP3 (e.g., HgMn) star. 

The position of HR\,62 in the log\,$g$--$T_\mathrm{eff}$ diagram (see Fig. \ref{fig2}) shows that the star has just left the main-sequence. If we trace its evolutionary track back to the main sequence, we reach the domain where the CP4 stars are. We may assume that the star could have been once of CP4 type (e.g. He-weak PGa) and now may have evolved into CP3 type. A more detailed analysis with a higher resolution spectrum of HR\,62 at shorter wavelengths ($\lambda$ < 4380 \AA{}) is needed to look for the Hg and Ga lines. HR\,62 may be a instance of a B-type CP giant migrating from CP4 to CP3 type.

\bibliography{Kilicoglu_T}  % For BibTex

\begin{thebibliography}{}
\expandafter\ifx\csname natexlab\endcsname\relax\def\natexlab#1{#1}\fi
\expandafter\ifx\csname url\endcsname\relax
  \def\url#1{\texttt{#1}}\fi
\expandafter\ifx\csname urlprefix\endcsname\relax\def\urlprefix{URL }\fi
\providecommand{\eprint}[2][]{\url{#2}}

\bibitem[{{Bessell} et~al.(1998){Bessell}, {Castelli}, \&
  {Plez}}]{besselletal98}
{Bessell}, M.~S., {Castelli}, F., \& {Plez}, B. 1998, \aap, 333, 231

\bibitem[{{Bressan} et~al.(2012){Bressan}, {Marigo}, {Girardi}, {Salasnich},
  {Dal Cero}, {Rubele}, \& {Nanni}}]{bressanetal12}
{Bressan}, A., {Marigo}, P., {Girardi}, L., {Salasnich}, B., {Dal Cero}, C.,
  {Rubele}, S., \& {Nanni}, A. 2012, \mnras, 427, 127. \eprint{1208.4498}

\bibitem[{{Flower}(1996)}]{flower96}
{Flower}, P.~J. 1996, \apj, 469, 355

\bibitem[{{Green} et~al.(2018){Green}, {Schlafly}, {Finkbeiner}, {Rix},
  {Martin}, {Burgett}, {Draper}, {Flewelling}, {Hodapp}, {Kaiser}, {Kudritzki},
  {Magnier}, {Metcalfe}, {Tonry}, {Wainscoat}, \& {Waters}}]{greenetal18}
{Green}, G.~M., {Schlafly}, E.~F., {Finkbeiner}, D., {Rix}, H.-W., {Martin},
  N., {Burgett}, W., {Draper}, P.~W., {Flewelling}, H., {Hodapp}, K., {Kaiser},
  N., {Kudritzki}, R.-P., {Magnier}, E.~A., {Metcalfe}, N., {Tonry}, J.~L.,
  {Wainscoat}, R., \& {Waters}, C. 2018, \mnras, 478, 651. \eprint{1801.03555}

\bibitem[{{Grevesse} \& {Sauval}(1998)}]{grevessesauval98}
{Grevesse}, N., \& {Sauval}, A.~J. 1998, \ssr, 85, 161

\bibitem[{{Hubeny} \& {Lanz}(1995)}]{hubenylanz95}
{Hubeny}, I., \& {Lanz}, T. 1995, \apj, 439, 875

\bibitem[{{K{\i}l{\i}{\c c}o{\u g}lu} \& {Monier}(2018)}]{kilicoglu18}
{K{\i}l{\i}{\c c}o{\u g}lu}, T., \& {Monier}, R. 2018, ArXiv e-prints.
  \eprint{1810.08444}

\bibitem[{{Kurucz}(2005)}]{kurucz05}
{Kurucz}, R.~L. 2005, Memorie della Societa Astronomica Italiana Supplementi,
  8, 14

\bibitem[{{Markwardt}(2009)}]{markwardt09}
{Markwardt}, C.~B. 2009, in Astronomical Data Analysis Software and Systems
  XVIII, edited by D.~A. {Bohlender}, D.~{Durand}, \& P.~{Dowler}, vol. 411 of
  Astronomical Society of the Pacific Conference Series, 251.
  \eprint{0902.2850}

\bibitem[{{Michaud} et~al.(2015){Michaud}, {Alecian}, \&
  {Richer}}]{michaudetal15}
{Michaud}, G., {Alecian}, G., \& {Richer}, J. 2015, {Atomic Diffusion in Stars,
  Astronomy and Astrophysics Library, ISBN 978-3-319-19853-8.} (Springer
  International Publishing Switzerland, 2015.)

\bibitem[{{Sbordone} et~al.(2004){Sbordone}, {Bonifacio}, {Castelli}, \&
  {Kurucz}}]{sbordoneetal04}
{Sbordone}, L., {Bonifacio}, P., {Castelli}, F., \& {Kurucz}, R.~L. 2004,
  Memorie della Societa Astronomica Italiana Supplementi, 5, 93.
  \eprint{astro-ph/0406268}

\end{thebibliography}

\end{document}